\documentclass[letterpaper]{article} 
\usepackage[preprint]{aaai2027}  
\usepackage[hyphens]{url}  
\usepackage{graphicx} 
\usepackage{enumitem}
\usepackage{natbib}  
\usepackage{caption} 
\usepackage{algorithm}
\usepackage{algorithmic}
\usepackage{amsmath}
\usepackage{amssymb}
\usepackage{newfloat}
\usepackage{listings}
\DeclareCaptionStyle{ruled}{labelfont=normalfont,labelsep=colon,strut=off} 
\floatstyle{ruled}
\newfloat{listing}{tb}{lst}{}
\floatname{listing}{Listing}

\usepackage{booktabs}

\title{BRA-Audit: Budgeted Runtime Auditing for LLM Multi-Agent Systems via Cumulative-Exposure Audit-Point Placement}
\author{
    Kaixiang Wang,
    Yidan Lin,
    Jiong Lou,
    Jie Li
}
\affiliations{
    
}

\begin{document}

\maketitle

\begin{abstract}

LLM-based multi-agent systems (LLM-MAS) solve complex tasks through specialized collaboration, but inter-agent dependencies can propagate hallucinated or malicious outputs into system-level failures. Auditor agents mitigate these risks, yet existing strategies face an efficiency dilemma: end-only auditing reviews long trajectories and final outputs, potentially weakening audit effectiveness and enlarging rollback scope, while auditing every agent each round improves detection and localization at high token cost. How can guard performance be preserved while minimizing token cost?
To address this problem, we propose BRA-Audit, a budget-aware runtime auditing framework that models MAS execution as a dynamic dependency graph and formulates audit scheduling as audit-point placement under a fixed audit-call budget to minimize cumulative unchecked exposure. Its greedy scheduler prioritizes influential and long-unaudited regions, while trusted audit points enable localized recovery. Across structured coordination, complex reasoning, and open-ended tasks, BRA-Audit restores performance close to the clean setting, remains competitive with heavy guard methods and reduces end-to-end token consumption by \(17.2\%\)--\(40.6\%\). 
\end{abstract}

\section{Introduction}
LLM-based multi-agent systems (LLM-MAS) coordinate specialized agents through explicit roles, communication, and shared workflows~\cite{hong2024metagpt,qian2024chatdev,chan2023chateval}. Recent systems demonstrate that MAS can support generalist task execution, diverse coordination protocols, multi-agent reflection, and layered response aggregation~\cite{fourney2024magenticonegeneralistmultiagentsolving,zhu2025multiagentbenchevaluatingcollaborationcompetition,wang2024mixtureofagentsenhanceslargelanguage}. Other studies optimize communication graphs and architecture selection to remove redundant exchanges and adapt computation to task difficulty~\cite{zhang2024cutcrapeconomicalcommunication,li2024survey,guo2024large}.
These advances suggest that LLM-MAS can combine task decomposition, heterogeneous expertise, tool use, and multiple reasoning paths, making them promising for long-horizon and complex open-ended tasks~\cite{tran2025collaboration,yan2026selftalk,li2023camel}.

However, multi-agent coordination also expands the failure surface. Hallucinated or malicious messages can be reused as context, alter downstream decisions, and propagate across agents~\cite{lee2024promptinfection,ju2024flooding,he2025redteaming}. Recent attacks can manipulate exchanged messages, hijack execution control, leak system information, or conceal malicious intent~\cite{triedman2025maliciouscode,wang2025masleak,xie2025mole}. Because agents repeatedly consume one another's outputs, a local error may evolve into a system-level failure, while its origin becomes difficult to identify in long and tightly coupled trajectories~\cite{xia2026agentlocate,zheng2025integrity}.

\begin{figure}[t]
    \centering
    \includegraphics[width=\linewidth]{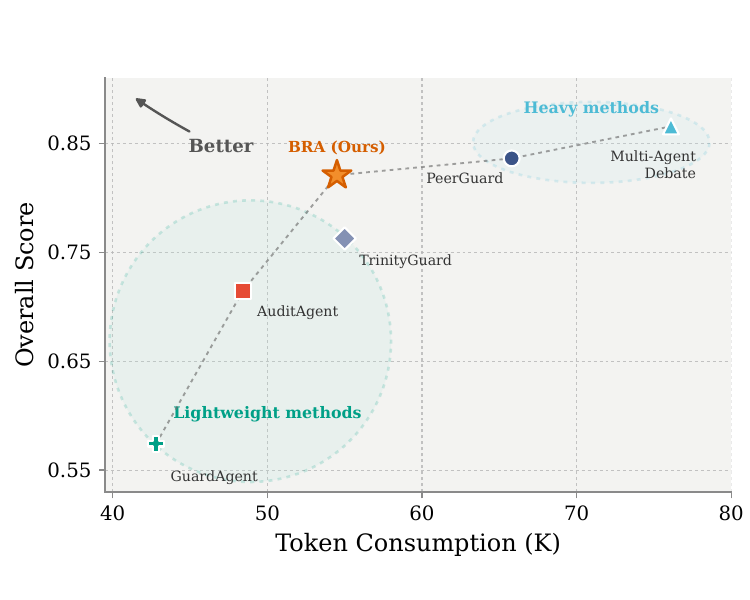}
    \caption{The performance-cost trade-off of BRA-Audit against other methods under attack in AgentsNet benchmark.}
    \label{fig:pareto}
\end{figure}

Past safeguards detect unreliable behavior through self-reflection or graph-based anomaly detection. However, self-reflection may reproduce the generator's failure modes, while GNN-based detectors often rely on labeled attacks and generalize poorly~\cite{shinn2023reflexion,renze2024selfreflection,wang2025gsafeguard,miao2025blindguard}. Dedicated auditor agents have therefore become a common alternative, providing stronger and more transferable oversight through runtime inspection and cross-agent reasoning~\cite{xiang2024guardagent,wang2026trinityguard,fan2025peerguard,du2024multiagentdebate}. Nevertheless, existing studies mainly improve \emph{how} auditing is performed, with limited attention to \emph{when} and \emph{where} auditors should be invoked. End-only auditing examines the full execution trajectory and final result after task completion. Its long, dependency-dense context can obscure failure sources, weaken audit effectiveness, delay intervention, and increase rollback costs~\cite{liu-etal-2024-lost,yang2026agentauditor}. In contrast, interaction-level auditing, as exemplified by PeerGuard~\cite{fan2025peerguard}, inspects agents at each round and enables earlier detection and finer localization. However, repeated auditor calls incur substantial token costs and waste computation on benign interactions~\cite{du2024multiagentdebate}. This raises a significant question: \emph{how can we determine when and where to audit while preserving guard performance and minimizing token cost?}

To answer this  question, we propose BRA-Audit, a budget-aware runtime
framework that directs limited audit resources to critical points in LLM-MAS execution. BRA models inter-agent communication and execution dependencies as a dynamic graph, where nodes represent runtime outputs and edges capture information flow. It then formulates audit scheduling as an audit-point placement problem that minimizes cumulative unchecked exposure under a fixed budget. The scheduler combines audit gaps with topological influence, prioritizing regions that have remained unchecked and can strongly affect downstream computation. Each verified audit point becomes a trusted audit point in this round; when unsafe behavior is detected, BRA contains the affected region and rolls back only the necessary subtrajectory. By integrating topology-aware scheduling, bounded auditing, and localized recovery, BRA maintains effective runtime oversight while reducing redundant checks and re-execution. It therefore achieves an empirical trade-off between audit performance and token consumption.

Further experimental results show that: across MAS collaboration environments and complex open-ended tasks, BRA-Audit restores task performance close to clean baselines while remaining competitive with stronger auditing baselines. It reduces
token consumption by \(17.2\%\)--\(40.6\%\) compared with heavy guard like PeerGuard and Multi-Agent Debate, demonstrating a favorable trade-off between defense effectiveness and token cost (as shown in Fig.~\ref{fig:pareto}).

Our contributions are as follows:
\begin{itemize}[leftmargin=2em]
    \item To determine when and where to audit an LLM-MAS while minimizing cost without compromising audit effectiveness, we formulate budget-constrained auditing as an audit-point placement problem over dynamic execution graphs.


    \item We propose BRA-Audit, an online budgeted audit-point placement framework in LLM-MAS that sequentially selects audit points by marginal cumulative-exposure reduction. It combines audit gaps and downstream dependency reachability to weight unchecked exposure, while verified audit points bound audit contexts and localize rollback.
    

    \item We conduct extensive experiments in MAS collaboration environments and complex open-ended tasks. BRA-Audit achieves competitive recovery performance while reducing token consumption by \(17.2\%\)--\(40.6\%\) compared with strong auditing baselines, demonstrating a favorable empirical Pareto trade-off.
\end{itemize}

\section{Related Work}

\paragraph{Attacks and Failures in LLM-MAS.}LLM-based multi-agent systems are vulnerable because agents exchange and reuse intermediate outputs, allowing local errors to escalate into system-level failures. Systematic analysis identifies failures in system design, inter-agent alignment, and task verification, revealing collaboration-specific risks beyond isolated agents~\cite{cemri2025why}. Adversaries can exploit the same dependencies: malicious instructions can self-propagate across agents~\cite{lee2024promptinfection}, manipulated knowledge can spread through conversations and persist in shared retrieval memories~\cite{ju2024flooding}, and communication attacks can intercept or modify inter-agent messages without compromising individual agents~\cite{he2025redteaming}. More severe control-flow hijacking can invoke unsafe tools, exfiltrate sensitive data, or execute malicious code~\cite{triedman2025maliciouscode}. Overall, communication, shared memory, and orchestration amplify benign errors and adversarial inputs, making early detection and containment essential for reliable LLM-MAS.

\paragraph{LLM Multi-Agent Safeguarding.}
Early guards use prompt-based self-reflection, but the reviewer may
share the generator's failure modes
~\cite{shinn2023reflexion,renze2024selfreflection}. G-Safeguard applies a graph neural
network to communication graphs to detect anomalous agents and prune
risky links, yet relies on labeled attacks and may generalize poorly to unseen threats
~\cite{wang2025gsafeguard,miao2025blindguard}. More reliable approaches
introduce LLM auditors. GuardAgent converts safety requirements
into executable checks, TrinityGuard coordinates runtime monitors,
PeerGuard uses mutual reasoning, and multi-agent debate performs
iterative cross-examination
~\cite{xiang2024guardagent,wang2026trinityguard,
fan2025peerguard,du2024multiagentdebate}. These methods improve
oversight, but their effectiveness and cost vary on complex tasks.

These safeguards methods either inspect trajectories too coarsely, which
weakens detection and enlarges the rollback scope, or audit every
interaction, which incurs substantial token and latency costs. We
therefore propose BRA-Audit, a budgeted audit-point placement framework
that preserves effective trajectory auditing while reducing redundant
checks.

\begin{figure*}[t]
    \centering
    \includegraphics[width=0.9\linewidth]{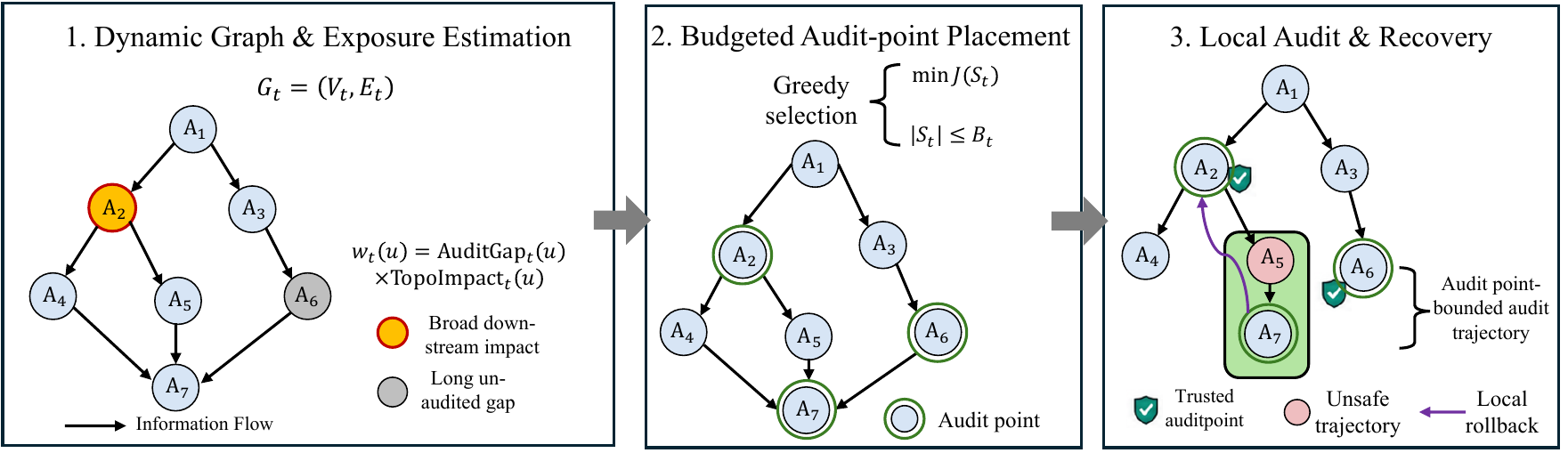}
    \caption{Overview of BRA-audit. }
    \label{fig:method}
\end{figure*}

\begin{algorithm}[t]
\caption{BRA: Exposure audit-point Scheduling}
\label{alg:bra_audit}
\begin{algorithmic}[1]
\REQUIRE Execution graph \(G_t=(V_t,E_t)\), candidate set
\(Q_{t+1}\subseteq V_t\), audit ratio \(\rho\), and last audit
records \(\mathrm{LastAudit}_t(\cdot)\)
\ENSURE Selected audit points \(S_{t+1}\) and updated audit
records \(\mathrm{LastAudit}_{t+1}(\cdot)\)

\STATE \(B_{t+1}\leftarrow
\left\lfloor \rho |Q_{t+1}| \right\rfloor\)
\STATE \(S_{t+1}\leftarrow \emptyset\)

\FOR{each node \(u\in V_t\)}
    \STATE \(\mathrm{AuditGap}_t(u)
    \leftarrow t-\mathrm{LastAudit}_t(a(u))+1\)
    \STATE \(\mathrm{TopoImpact}_t(u)
    \leftarrow 1+|\mathrm{Desc}_t(u)|\)
    \STATE \(w_t(u)\leftarrow
    \mathrm{AuditGap}_t(u)\cdot
    \mathrm{TopoImpact}_t(u)\)
\ENDFOR

\WHILE{\(|S_{t+1}|<B_{t+1}\) and
\(Q_{t+1}\setminus S_{t+1}\neq\emptyset\)}
    \FOR{each \(v\in Q_{t+1}\setminus S_{t+1}\)}
        \STATE \(\Delta(v\mid S_{t+1})
        \leftarrow J(S_{t+1})
        -J(S_{t+1}\cup\{v\})\)
    \ENDFOR
    \STATE \(v^\star\leftarrow
    \arg\max_{v\in Q_{t+1}\setminus S_{t+1}}
    \Delta(v\mid S_{t+1})\)
    \STATE \(S_{t+1}\leftarrow
    S_{t+1}\cup\{v^\star\}\)
\ENDWHILE

\STATE deploy \(S_{t+1}\) before round-\((t+1)\) execution

\FOR{each audit point \(s\in S_{t+1}\) in causal order}
    \STATE \(\mathcal{T}(s)\leftarrow
    \operatorname{Region}(s\mid S_{t+1})\)
    \STATE \(r\leftarrow
    \operatorname{Auditor}(\mathcal{T}(s))\)

    \IF{\(r=\mathrm{safe}\)}
        \STATE mark \(s\) as a trusted audit point for
        downstream execution in this round
    \ELSE
        \STATE block the affected downstream information flow
        \STATE rollback and recover the work products
        affected by \(\mathcal{T}(s)\)
        \STATE re-execute and re-audit \(\mathcal{T}(s)\)
        before continuing
    \ENDIF

    \STATE update
    \(\mathrm{LastAudit}_{t+1}(a(s))\)
\ENDFOR

\RETURN \(S_{t+1},
\mathrm{LastAudit}_{t+1}(\cdot)\)
\end{algorithmic}
\end{algorithm}

\section{System Model}
\label{system model}

This section defines the LLM-MAS execution model and the threat model considered in this work.

\subsection{MAS System}

We consider a graph-structured LLM multi-agent system (LLM-MAS) with a set of agents
\begin{equation}
\mathcal{A}=\{a_1,\ldots,a_n\}.
\end{equation}
Each agent has a predefined role and contributes to the task through its own interaction pattern, information sources, and tools. We represent runtime execution as a dynamic dependency graph:
\begin{equation}
G_t=(V_t,E_t).
\end{equation}
Each node \(v\in V_t\) represents a runtime object, such as an agent message, intermediate result, tool interaction, retrieved evidence, or aggregation result. It is associated with its producing agent \(a(v)\), timestamp \(t(v)\), and content \(m(v)\). A directed edge \((u,v)\in E_t\) indicates that \(v\) depends on \(u\) through communication, evidence usage, tool execution, or aggregation. A path in \(G_t\) therefore captures the propagation of information and potential errors.

Auditing is modeled as audit-point placement on the execution graph. An audit point is a selected node \(s\in V_t\) inspected by an auditor using the relevant context between the previous audit point and \(s\). If the trajectory is safe, \(s\) becomes a trusted audit point in this round. Otherwise, the system may block, sanitize, isolate, or roll back the affected trajectory.

\subsection{Threat Model}

Unsafe behavior may arise from compromised agents, hallucinations, tool errors, poisoned knowledge bases, corrupted websites, or misleading retrieved documents. Despite their different causes, these failures introduce false, unsafe, or misleading information into the execution graph.

We use \emph{malicious agents} as a unified abstraction for these cases. Let
\begin{equation}
\mathcal{M}\subseteq \mathcal{A},
\end{equation}
denote the set of malicious agents. For a node \(v\in V_t\), its content may be unsafe if it is produced by a malicious agent:
\begin{equation}
a(v)\in \mathcal{M}
\Longrightarrow
\Pr(y(v)=1)>0.
\end{equation}
Here, \(y(v)=1\) denotes unsafe content. The term \emph{malicious} does not require intentional behavior; it only indicates that an agent produces corrupted information that may mislead downstream agents.

A malicious agent may fabricate evidence, omit constraints, produce incorrect intermediate results, distort aggregation, or trigger unsafe tool usage. These outputs can propagate through dependency edges and affect later reasoning or the final result. We therefore study how runtime auditing detects and contains such propagated failures over the execution graph.

\section{Method}
This section formulates budgeted runtime auditing as an optimization problem and develops the mathematical model and scheduling algorithm used by BRA-Audit (Fig. ~\ref{fig:method} shows overview of BRA-audit).

\subsection{Problem Formulation and Optimization Objective}

LLM-based auditors can inspect intermediate MAS outputs and identify hallucinated, inconsistent, or unsafe reasoning
steps~\citep{zheng2023judgingllmasajudgemtbenchchatbot,
gu2025surveyllmasajudge}. As discussed in the introduction, final-only auditing requires processing a long execution trajectory and may cause large rollback, whereas auditing every interaction requires frequent auditor calls. We therefore study how to allocate a limited audit budget during MAS execution.

A successful audit verifies the audit point-bounded dependency
region reaching the inspected node and establishes the node as
a trusted audit point in this round for subsequent execution. The size of this
region determines both the context inspected by the auditor and
the execution scope that may need to be rolled back. Audit-point
placement is therefore important: leaving a node with broad
downstream reachability unchecked can repeatedly carry its
exposure into later audit contexts, while a long audit gap
indicates stale trust evidence for the corresponding agent or
region. Moreover, audit point decisions are coupled because each
new audit point reshapes downstream audit regions and changes the
value of subsequent audit points.

Given a candidate set \(Q_t\subseteq V_t\) and an audit-call
budget \(B_t\), BRA-Audit selects before execution an audit point
set
\begin{equation}
S_t\subseteq Q_t,
\qquad
|S_t|\leq B_t.
\end{equation}
Each selected audit point triggers one auditor call over its
audit-point bounded dependency region. The scheduler therefore
seeks placements that keep audit regions compact while
prioritizing influential nodes and agents that have remained
unaudited for many rounds.

We formulate audit point placement as minimizing cumulative
unchecked exposure:
\begin{equation}
\min_{S_t\subseteq Q_t} J(S_t)
\quad
\mathrm{s.t.}
\quad
|S_t|\leq B_t,
\end{equation}
where \(J(S_t)\) measures the total cumulative exposure
remaining under audit-point placement \(S_t\). By preventing
unchecked information from being repeatedly inherited by
downstream contexts, this objective jointly reduces audit
context size and delayed-detection exposure. Here, \(B_t\)
controls the number of auditor calls, while total token
cost is evaluated end to end. We define \(J(S_t)\) and
its components in the next subsection.

\subsection{Analysis and Solution}

We now analyze the cumulative-exposure objective and derive the audit-point scheduling rule used by BRA-Audit. Algorithm~\ref{alg:bra_audit} summarizes the main procedure of BRA-Audit. The key observation is that an unchecked node does not only affect its own audit context. Its exposure can continue to appear in downstream dependency contexts until the corresponding information flow reaches an audit point. Audit scheduling therefore considers how an audit point jointly reduces exposure in multiple downstream contexts.

Since runtime objects are ordered by their generation times and dependency edges follow causal information flow, repeated agent interactions are unrolled across rounds, and the runtime dependency graph \(G_t\) is represented as a directed acyclic graph.

For each node \(u\in V_t\), let \(a(u)\) denote its producing agent, and let \(\tau_t(a)\) be the latest completed audit round of agent \(a\) available at scheduling time \(t\). We define 
\begin{equation} \begin{aligned} \mathrm{Audit Gap}(u) &= (t+1)-\tau_t(a(u)),\\ \mathrm{TopoImpact}_t(u) &= \left|\{u\}\cup\mathrm{Desc}_t(u)\right|,\\ w_{t+1\mid t}(u) &= \mathrm{AuditGap}(u)\, \mathrm{TopoImpact}_t(u), 
\end{aligned} 
\end{equation} 

where \(\mathrm{Desc}_t(u)\) denotes the strict descendants of \(u\).
We consider verification age because risk can accumulate while an
agent's outputs remain unchecked, and downstream reach because errors
at influential nodes can affect more subsequent computations and nodes with broad downstream reachability also tend to play more critical roles in the task. Including
\(u\) itself ensures that even sink nodes retain nonzero local impact.
Accordingly, the first term measures the verification age when the
scheduled audit is executed, while the second measures the closed
downstream region potentially affected by \(u\). Their product serves
as a temporal--topological exposure measure, prioritizing nodes with
stale verification and broad downstream reach using runtime audit
records and graph structure.

Let \(\mathrm{Anc}_t(v)\) denote the ancestors of \(v\).
We write \(u\leadsto_{S_t}v\) if there exists a directed path
\(p:u\leadsto v\) such that
\((V(p)\setminus\{v\})\cap S_t=\varnothing\).
Given an audit point set \(S_t\), the unchecked audit context
reaching \(v\) is
\begin{equation}
\mathrm{Ctx}(v\mid S_t)
=
\left\{
u\in\mathrm{Anc}_t(v)\cup\{v\}
:
u\leadsto_{S_t}v
\right\}.
\end{equation}
Equivalently, it is obtained by traversing backward from \(v\)
and stopping each branch at the first selected audit point.
This definition preserves unresolved branches in a DAG: if
any path from \(u\) to \(v\) bypasses the selected audit points,
\(u\) remains in the audit context. Thus, an audit point shortens
only the downstream contexts whose information flow passes
through it.

All audit points are selected before execution and triggered
in causal order. Once an audit point passes the audit, its
output becomes trusted input for downstream execution; an
unsafe region is blocked and recovered before execution
continues.

The cumulative exposure of \(v\) and the total exposure are
\begin{equation}
\begin{aligned}
C(v\mid S_t)
&=
\sum_{u\in\mathrm{Ctx}(v\mid S_t)}w_t(u),\\
J(S_t)
&=
\sum_{v\in V_t}C(v\mid S_t).
\end{aligned}
\end{equation}
A high-exposure node can repeatedly contribute to downstream
contexts until bounded by an audit point, increasing the audit
burden and the cost of delayed detection. By summing over all
nodes, \(J(S_t)\) captures both the size of local audit contexts
and the repeated inheritance of unchecked exposure by
downstream computation.

Directly minimizing \(J(S_t)\) is combinatorial, and
audit point decisions are coupled because each selected
audit point changes downstream contexts and the value of
later selections. BRA-Audit therefore greedily selects
audit points by marginal gain.

\begin{table*}[t]
\centering
\caption{Performance and token cost on AgentsNet.
Token consumption is reported in thousands (K).}
\label{tab:agentsnet_main}

\small
\setlength{\tabcolsep}{6pt}
\renewcommand{\arraystretch}{1.10}

\resizebox{0.9\linewidth}{!}{%
\begin{tabular}{@{}l*{12}{c}@{}}
\toprule
& \multicolumn{2}{c}{\textbf{Consensus}}
& \multicolumn{2}{c}{\textbf{Leader}}
& \multicolumn{2}{c}{\textbf{Matching}}
& \multicolumn{2}{c}{\textbf{Coloring}}
& \multicolumn{2}{c}{\textbf{VertexCover}}
& \multicolumn{2}{c}{\textbf{Overall}} \\
\cmidrule(lr){2-3}
\cmidrule(lr){4-5}
\cmidrule(lr){6-7}
\cmidrule(lr){8-9}
\cmidrule(lr){10-11}
\cmidrule(lr){12-13}

\textbf{Method}
& \textbf{Score} & \textbf{Tok.}
& \textbf{Score} & \textbf{Tok.}
& \textbf{Score} & \textbf{Tok.}
& \textbf{Score} & \textbf{Tok.}
& \textbf{Score} & \textbf{Tok.}
& \textbf{Score} & \textbf{Tok.} \\
\midrule

Clean
& 0.9630 & 28.1
& 0.5926 & 29.9
& 0.6944 & 25.7
& 0.9967 & 31.1
& 0.8342 & 18.6
& 0.8162 & 26.7 \\

Attack
& 0.0741 & 42.2
& 0.1481 & 43.1
& 0.4815 & 35.7
& 0.8324 & 47.5
& 0.6676 & 29.0
& 0.4407 & 39.5 \\

\midrule

PeerGuard
& 0.9630 & 69.3
& 0.5556 & 67.5
& 0.8056 & 68.1
& 0.9716 & 70.4
& 0.8852 & 53.6
& 0.8362 & 65.8 \\

AuditAgent
& 0.8148 & 48.6
& 0.4444 & 50.2
& 0.6250 & 49.1
& 0.9013 & 53.9
& 0.7854 & 40.3
& 0.7142 & 48.4 \\

GuardAgent
& 0.4074 & 44.7
& 0.3704 & 46.5
& 0.5370 & 39.3
& 0.8548 & 50.2
& 0.7014 & 33.4
& 0.5742 & 42.8 \\

TrinityGuard
& 0.8889 & 58.0
& 0.4815 & 55.6
& 0.7385 & 51.8
& 0.8892 & 64.8
& 0.8156 & 44.9
& 0.7627 & 55.0 \\

Multi-Agent Debate
& 1.0000 & 75.4
& 0.6296 & 74.5
& 0.8342 & 77.3
& 0.9967 & 81.6
& 0.8675 & 71.5
& 0.8656 & 76.1 \\

\midrule

\textbf{BRA-Audit}
& 0.9630 & 57.6
& 0.5185 & 57.0
& 0.7917 & 52.4
& 0.9967 & 61.7
& 0.8342 & 43.8
& 0.8208 & 54.5 \\

\bottomrule
\end{tabular}%
}
\end{table*}

For a candidate \(v\in V_t\setminus S_t\), let
\(\mathrm{Cut}_v(z\mid S_t)\) denote the exposure removed
from the context of \(z\):
\begin{equation}
\begin{aligned}
\mathrm{Cut}_v(z\mid S_t)
&=
\mathrm{Ctx}(z\mid S_t)\\
&\quad\setminus
\mathrm{Ctx}(z\mid S_t\cup\{v\}).
\end{aligned}
\end{equation}
The marginal gain of \(v\) is
\begin{equation}
\begin{aligned}
\Delta(v\mid S_t)
&=
J(S_t)-J(S_t\cup\{v\})\\
&=
\sum_{z\in V_t}
\sum_{u\in\mathrm{Cut}_v(z\mid S_t)}
w_t(u).
\end{aligned}
\end{equation}
This quantity measures the exposure removed from all
dependency contexts by adding \(v\). It is large when \(v\)
separates a high-exposure region from many downstream
contexts. Since marginal gains are recomputed after each
selection, the scheduler accounts for overlapping effects
among audit points rather than scoring them independently.
For a chain, the expression reduces to the accumulated prefix
exposure multiplied by the number of affected downstream
contexts, as illustrated in the technical appendix.

At each step, BRA-Audit selects
\begin{equation}
v^\star
=
\arg\max_{v\in V_t\setminus S_t}
\Delta(v\mid S_t),
\end{equation}
adds \(v^\star\) to \(S_t\), and updates the affected contexts
until the audit budget is exhausted.

\begin{table*}[t]
\centering
\caption{Performance and token cost on BBH and Multi-agent-bench-rasearch.}
\label{tab:BBH}
\resizebox{0.9\linewidth}{!}{%
\begin{tabular}{cccccccccc}
\hline
                            & \multicolumn{3}{c}{\textbf{BBH}}                  & \multicolumn{6}{c}{\textbf{MultiAgentBench-Research}}                                                                 \\ \hline
\textbf{Method}             & \textbf{Acc} & \textbf{recovery} & \textbf{token} & \textbf{innovation} & \textbf{safety} & \textbf{feasibility} & \textbf{Score} & \textbf{recovery} & \textbf{token(K)} \\ \hline
\textbf{clean}              & 0.9200       & --                & 5355           & 5.00                & 4.62            & 3.83                 & 0.897          & --                & 23.0              \\
\textbf{attack}             & 0.3933       & --                & 6778           & 2.28                & 1.88            & 1.50                 & 0.377          & --                & 29.1              \\
\textbf{PeerGuard}          & 0.8867       & 0.4933            & 10136          & 4.96                & 4.80            & 3.86                 & 0.908          & 0.531             & 47.3              \\
\textbf{Auditagent}         & 0.8067       & 0.4133            & 7459           & 3.88                & 3.92            & 3.14                 & 0.730          & 0.352             & 32.4              \\
\textbf{Guardagent}         & 0.5200       & 0.1267            & 7318           & 3.20                & 2.74            & 2.88                 & 0.588          & 0.211             & 30.8              \\
\textbf{Trinityguard}       & 0.8600       & 0.4667            & 8324           & 4.32                & 4.16            & 3.64                 & 0.808          & 0.431             & 37.9              \\
\textbf{Multi-Agent Debate} & 0.9333       & 0.54              & 13763          & 4.98                & 4.86            & 4.02                 & 0.925          & 0.548             & 60.5              \\ \hline
\textbf{BRA}                & 0.8600       & 0.4667            & 8175           & 5.00                & 4.74            & 3.78                 & 0.902          & 0.525             & 38.5              \\ \hline
\end{tabular}
}
\end{table*}

\subsection{Online audit-point Placement and Audit Execution}

BRA-Audit operates in an online manner. At the end of each MAS interaction round \(t\), the system updates the dependency graph \(G_t=(V_t,E_t)\). Before the next round starts, the scheduler uses this graph to place a limited number of audit points. Let \(Q_{t+1}\subseteq V_t\) denote the candidate nodes for the next audit round. Given an audit ratio \(\rho\in(0,1]\), which corresponds to auditing \(k\%\) of the candidates, the audit budget and audit point set are defined as
\begin{equation}
B_{t+1}=\left\lfloor \rho |Q_{t+1}| \right\rfloor,
\qquad
S_{t+1}\approx
\arg\min_{S\subseteq Q_{t+1},\, |S|\leq B_{t+1}} J(S).
\end{equation}
Here, \(J(S)\) is the cumulative unaudited exposure objective defined above. In practice, BRA-Audit selects audit points that provide the largest reduction of \(J(S)\) until the budget is used. In the final round, the agent producing the final result is mandatorily selected and counted toward the budget.


After the audit-point set \(S_{t+1}\) is selected, each
audit point is assigned to an auditor. For an audit point
\(s\in S_{t+1}\), the auditor inspects the audit-point bounded
dependency region accumulated along all incoming branches:
\begin{equation}
\mathcal{T}(s\mid S_{t+1})
=
\operatorname{Region}(s\mid S_{t+1}).
\end{equation}

Here, \(\operatorname{Region}(s\mid S_{t+1})\) is obtained by
traversing backward from \(s\) and stopping each branch at
the first upstream audit point in \(S_{t+1}\). The audit point
outputs are treated as trusted inputs, while the subsequent
inputs, outputs, messages, and dependency steps form the local
audit region.

If the auditor judges \(\mathcal{T}(s\mid S_{t+1})\) as safe,
\(s\) becomes a trusted audit point; otherwise, the system
rolls back the work products affected by this region. By
partitioning execution into audit-point bounded regions,
BRA-Audit shortens audit contexts and localizes rollback.
Because the auditor examines the entire region rather than
only its endpoint, an unsafe agent can still be detected even
when it is not selected as an audit point, provided that its
inputs or outputs appear in the region of an downstream
audit point.

\begin{table*}[t]
\centering
\caption{Backbone model ablation on AgentsNet.
Token consumption is reported in thousands (K).}
\label{tab:model}

\footnotesize
\setlength{\tabcolsep}{6pt}
\renewcommand{\arraystretch}{1.10}

\resizebox{0.9\linewidth}{!}{%
\begin{tabular}{@{}l*{12}{c}@{}}
\toprule
& \multicolumn{2}{c}{\textbf{Consensus}}
& \multicolumn{2}{c}{\textbf{Leader}}
& \multicolumn{2}{c}{\textbf{Matching}}
& \multicolumn{2}{c}{\textbf{Coloring}}
& \multicolumn{2}{c}{\textbf{VertexCover}}
& \multicolumn{2}{c}{\textbf{Overall}} \\
\cmidrule(lr){2-3}
\cmidrule(lr){4-5}
\cmidrule(lr){6-7}
\cmidrule(lr){8-9}
\cmidrule(lr){10-11}
\cmidrule(lr){12-13}

\textbf{Model}
& \textbf{Score} & \textbf{Tok.}
& \textbf{Score} & \textbf{Tok.}
& \textbf{Score} & \textbf{Tok.}
& \textbf{Score} & \textbf{Tok.}
& \textbf{Score} & \textbf{Tok.}
& \textbf{Score} & \textbf{Tok.} \\
\midrule

Qwen
& 0.9630 & 57.6
& 0.5185 & 57.0
& 0.7917 & 52.4
& 0.9967 & 61.7
& 0.8342 & 43.8
& 0.8208 & 54.5 \\

DeepSeekV4-Flash
& 0.9630 & 51.0
& 0.5556 & 72.9
& 0.6435 & 64.2
& 0.9463 & 51.3
& 0.8250 & 44.6
& 0.7867 & 56.8 \\

MiniMax-2.7
& 0.9259 & 55.8
& 0.5185 & 63.4
& 0.7592 & 51.3
& 0.9177 & 55.4
& 0.7935 & 35.6
& 0.7830 & 52.3 \\

GLM-5.1
& 0.9259 & 62.9
& 0.5926 & 63.2
& 0.7314 & 56.8
& 0.9446 & 61.0
& 0.8250 & 53.1
& 0.8039 & 59.4 \\

GPT-5
& 0.9630 & 78.5
& 0.6296 & 79.8
& 0.8333 & 81.7
& 0.9534 & 75.9
& 0.8956 & 65.1
& 0.8550 & 76.2 \\

Gemini3-Flash
& 0.8889 & 48.7
& 0.5556 & 51.5
& 0.7963 & 50.1
& 0.9265 & 53.7
& 0.8342 & 40.2
& 0.8003 & 48.8 \\

\bottomrule
\end{tabular}%
}
\end{table*}

\begin{table*}[t]
\centering
\caption{Agent-count and topology ablations. Each entry reports
Score / Avg.\ Token (K).}
\label{tab:scale_topology_ablation}

\small
\renewcommand{\arraystretch}{1.15}

\begin{minipage}[t]{0.48\textwidth}
\centering
\textbf{(a) Number of Agents}

\medskip
\begin{tabular*}{\linewidth}
{@{\extracolsep{\fill}}cccc@{}}
\toprule
\textbf{\#Agents}
& \textbf{Clean}
& \textbf{Attack}
& \textbf{BRA-Audit} \\
\midrule
4
& 0.859 / 8.3
& 0.363 / 14.1
& \textbf{0.837 / 18.9} \\

8
& 0.827 / 21.3
& 0.520 / 33.4
& \textbf{0.847 / 41.9} \\

16
& 0.763 / 50.4
& 0.439 / 71.0
& \textbf{0.778 / 102.8} \\
\midrule
Overall
& 0.816 / 26.7
& 0.441 / 39.5
& \textbf{0.821 / 54.5} \\
\bottomrule
\end{tabular*}
\end{minipage}
\hfill
\begin{minipage}[t]{0.48\textwidth}
\centering
\textbf{(b) Graph Topology}

\medskip
\begin{tabular*}{\linewidth}
{@{\extracolsep{\fill}}cccc@{}}
\toprule
\textbf{Topology}
& \textbf{Clean}
& \textbf{Attack}
& \textbf{BRA-Audit} \\
\midrule
WS
& 0.803 / 26.2
& 0.451 / 38.5
& \textbf{0.846 / 55.6} \\

BA
& 0.835 / 25.2
& 0.415 / 38.1
& \textbf{0.813 / 47.8} \\

DT
& 0.811 / 28.7
& 0.456 / 42.0
& \textbf{0.804 / 60.1} \\
\midrule
Overall
& 0.816 / 26.7
& 0.441 / 39.5
& \textbf{0.821 / 54.5} \\
\bottomrule
\end{tabular*}
\end{minipage}

\end{table*}
\section{Experiment}
We evaluate BRA-Audit against recent LLM-agent auditing and multi-agent defense methods, focusing on defense effectiveness, audit cost, and robustness. Our experiments are designed to answer three research questions:

\textbf{RQ1: Defense-Cost Trade-off.} 
Can BRA-Audit achieve effective auditing while keeping the audit overhead low?

\textbf{RQ2: Robustness Across Settings.} 
How stable is BRA-Audit under different graph topologies, node importance distributions, attack locations, and backbone auditor models?

\textbf{RQ3: Budget and Component Analysis.} 
How do audit budget, audit point placement strategy, and exposure-weight components affect the overall performance of BRA-Audit?

\subsection{Experimental Setup.}

\paragraph{Dataset.} We evaluate BRA-Audit on three complementary settings. First, we use AgentsNet~\citep{grötschla2025agentsnetcoordinationcollaborativereasoning}, a multi-agent coordination benchmark with five graph-theoretic task and configurable communication topologies and node counts. This setting tests auditing under controlled multi-round agent communication. Second, we use BIG-Bench Hard (BBH)~\citep{suzgun2022challengingbigbenchtaskschainofthought} to evaluate complex logical reasoning; we randomly sample 150 questions and convert each instance into a multi-agent reasoning workflow. Third, we use the Research scenario from MultiAgentBench~\citep{zhu2025multiagentbenchevaluatingcollaborationcompetition}, an open-ended multi-agent task where agents collaborate to generate research ideas.

\paragraph{Settings.}
We implement all multi-agent systems with AutoGen~\citep{wu2023autogen}, and instantiate agent networks with different sizes and communication topologies. In each setting, 20\% of the agents is designated as malicious and is instructed to inject incorrect claims, misleading evidence, or corrupted 
intermediate results into the collaboration process to make the task fail \textbf{(detailed attack recovery protocol is provided in technical supplement)}. We mainly use Qwen3-235B-A22B-Instruct-2507 model. We further evaluate different backbone models, graph topologies, numbers of agents, and audit budgets \(\rho\). The main metrics are task score and total E2E token consumption, which can measure defense effectiveness and cost. Each configuration is repeated three times, with the standard deviation kept below \(2\%\).

\subsection{Main Results.}

We compare BRA-Audit with five representative LLM-based methods for runtime auditing or reliability enhancement. PeerGuard performs peer-based inspection during agent interactions~\citep{fan2025peerguard}. AuditAgent is a round-level auditing baseline adapted from AgentAuditor~\citep{yang2026agentauditor}. GuardAgent converts scenario-specific guard requirements into executable checks~\citep{xiang2024guardagent}. TrinityGuard uses preassigned runtime monitors to inspect MAS execution~\citep{wang2026trinityguard}. Multi-Agent Debate improves output reliability through iterative discussion among agents~\citep{du2024multiagentdebate}. All methods use the same attack settings and backbone models.

The attack results confirm that LLM-MAS is vulnerable to a small number
of corrupted agents. Even one or two malicious or hallucinating agents
can propagate faulty outputs through communication and aggregation,
substantially degrading task performance. On AgentsNet, the overall
score falls from \(0.8162\) in the clean setting to \(0.4407\) under
attack, a relative drop of approximately \(46.0\%\). This supports auditing MAS collaboration is necessary.

On AgentsNet (Table~\ref{tab:agentsnet_main}), BRA-Audit restores the
overall score to \(0.8208\) with \(54.5\)K tokens. It outperforms
AuditAgent, GuardAgent, and TrinityGuard, and matches the clean results
on Consensus, Coloring, and VertexCover. Although PeerGuard and
Multi-Agent Debate achieve slightly higher scores, BRA-Audit uses
\(17.2\%\) and \(28.4\%\) fewer tokens, respectively.
A similar defense--cost trade-off appears on BBH and
MultiAgentBench-Research (Table~\ref{tab:BBH}). BRA-Audit improves BBH
accuracy from \(0.3933\) to \(0.8600\), using \(19.3\%\) fewer tokens
than PeerGuard and \(40.6\%\) fewer than Multi-Agent Debate. On
Research, it reaches \(0.902\) with \(38.5\)K tokens, close to both
baselines while reducing token usage by \(18.5\%\) and \(36.3\%\),
respectively. Overall, BRA-Audit balances defense effectiveness and
runtime cost across structured and open-ended tasks.

\subsection{Ablation Study}

\paragraph{Backbone Model Ablation.}
We evaluate BRA-Audit with six backbone models on AgentsNet benchmark to study its sensitivity to model capability(as shown in Table~\ref{tab:model}.) Overall scores range from \(0.7830\) to \(0.8550\), indicating stable behavior across all tested models. GPT-5 obtains the highest score of \(0.8550\), but also incurs the largest token cost at \(76.2\)K. Qwen offers a better performance--cost balance, reaching \(0.8208\) with \(54.5\)K tokens. Gemini-3 uses the fewest tokens (\(48.8\)K) while maintaining a score of \(0.8003\). GLM-5.1, DeepSeek-V4, and MiniMax-2.7 achieve \(0.8039\), \(0.7867\), and \(0.7830\), respectively. Performance differences are most visible on Leader and Matching. Overall, BRA-Audit remains effective across different backbones, while stronger models tend to improve task performance at higher token cost.

\begin{table}[t]
\centering
\caption{Audit-budget ablation with audit and rollback costs.
All token costs are reported in thousands (K).}
\label{tab:budget_ablation}

\footnotesize
\setlength{\tabcolsep}{2.6pt}
\renewcommand{\arraystretch}{1.10}

\resizebox{\columnwidth}{!}{%
\begin{tabular}{@{}l*{8}{c}@{}}
\toprule
& \multicolumn{4}{c}{\textbf{BBH}}
& \multicolumn{4}{c}{\textbf{Research}} \\
\cmidrule(lr){2-5}
\cmidrule(lr){6-9}

\textbf{\(\rho\)}
& \textbf{Score}
& \textbf{Audit}
& \textbf{Rollback}
& \textbf{Total}
& \textbf{Score}
& \textbf{Audit}
& \textbf{Rollback}
& \textbf{Total} \\
\midrule

\(0.2\)
& 0.760 & 0.74 & 2.84 & 6.90
& 0.757 & 2.24 & 13.14 & 33.2 \\

\(0.4\), Rand.
& 0.820 & 2.14 & 1.76 & 8.01
& 0.821 & 5.43 & 9.13 & 37.4 \\

\textbf{\(0.4\), BRA}
& \textbf{0.860} & 2.09 & 1.81 & 8.18
& \textbf{0.902} & 5.82 & 9.38 & 38.5 \\

\(0.6\)
& 0.873 & 3.91 & 1.35 & 9.57
& 0.906 & 9.11 & 7.17 & 42.9 \\

\(1.0\)
& 0.893 & 5.57 & 0.94 & 11.31
& 0.897 & 11.13 & 5.64 & 46.8 \\

\bottomrule
\end{tabular}%
}

\end{table}

\paragraph{Audit-Budget Ablation.}

We vary the audit ratio \(\rho\) on BBH and MultiAgentBench-Research to
study the defense--cost trade-off. On BBH, increasing \(\rho\) from
\(0.2\) to \(0.4\) improves the score from \(0.760\) to \(0.860\),
while total token consumption rises from \(6.9\) to \(8.18\).
Larger budgets of \(0.6\) and \(1.0\) further increase the score to
\(0.873\) and \(0.893\), but require \(9.57\) and \(11.31\)K
tokens. MultiAgentBench-Research shows a similar trend:
\(\rho=0.4\) achieves \(0.902\) with \(38.5\)K tokens, close to the
best score of \(0.906\) at \(\rho=0.6\). As \(\rho\) increases, audit
cost grows while rollback cost decreases, indicating that more frequent
auditing detects failures earlier and reduces re-execution. Under the
same budget \(\rho=0.4\), BRA-Audit outperforms random audit-point
placement by \(0.040\) on BBH and \(0.081\) on Research with comparable
token costs. Overall, \(\rho=0.4\) provides a favorable balance among
audit effectiveness, audit cost, and rollback overhead.

\paragraph{Node-Count Ablation.}
We evaluate BRA-Audit on AgentsNet graphs with 4, 8, and 16 agents
(Table~\ref{tab:scale_topology_ablation}a). It achieves scores of
\(0.8370\), \(0.8473\), and \(0.7781\), improving over the attacked
setting by \(0.4737\), \(0.3274\), and \(0.3391\), respectively.
Token consumption rises from \(18.9\)K to \(41.9\)K and \(102.8\)K as
larger graphs introduce longer communication paths and audit contexts.
BRA-Audit therefore remains effective as the system scales, although
larger networks require substantially more auditing resources.

\paragraph{Topology Ablation.}
We evaluate Watts--Strogatz, Barabási--Albert, and Delaunay graphs,
which represent clustered small-world, hub-dominated, and spatially
local communication structures, respectively
\citep{grötschla2025agentsnetcoordinationcollaborativereasoning}.
As shown in Table~\ref{tab:scale_topology_ablation}b, BRA-Audit obtains
scores of \(0.8458\), \(0.8127\), and \(0.8040\), compared with attacked
scores of \(0.4511\), \(0.4153\), and \(0.4558\). The small-world graph
achieves the best score, while the preferential-attachment graph uses
the fewest tokens. Overall, BRA-Audit remains effective across diverse
communication topologies.

\begin{table}[t]
\centering
\caption{False-positive results under different audit ratios.}
\label{tab:false_positive}
\footnotesize
\renewcommand{\arraystretch}{1.12}

\begin{tabular*}{\columnwidth}{
@{\extracolsep{\fill}}
c
cc
cc
@{}
}
\toprule
& \multicolumn{2}{c}{\textbf{BBH}}
& \multicolumn{2}{c}{\textbf{Research}} \\
\cmidrule(lr){2-3}
\cmidrule(lr){4-5}
\(\boldsymbol{\rho}\)
& \textbf{FP}
& \textbf{FPR}
& \textbf{FP}
& \textbf{FPR} \\
\midrule
\(0.2\) & 0  & \(0.00\%\)  & 0  & \(0.00\%\)  \\
\(0.4\) & 7  & \(4.67\%\)  & 6  & \(6.00\%\)  \\
\(1.0\) & 18 & \(12.00\%\) & 16 & \(16.00\%\) \\
\bottomrule
\end{tabular*}
\end{table}

\paragraph{False-Positive Analysis.}
We assess false positives on clean BBH and MultiAgentBench-Research at varying audit ratios (Table~\ref{tab:false_positive}). No FP occur at \(\rho=0.2\). At \(\rho=0.4\), FP rates are
\(4.67\%/6.00\%\) for BBH/Research, rising to
\(12.00\%/16.00\%\) under full auditing. Thus, excessive auditing can introduce unnecessary errors on benign trajectories. Most executions contain neither attacks nor severe hallucinations, so auditing every interaction incurs substantial redundant cost. Alongside task performance, BRA-Audit at \(\rho=0.4\) empirically
lies on a favorable Pareto frontier balancing defense effectiveness, FP risk, and token consumption.


\paragraph{Weight-Component Ablation.} We further ablate the two components of the exposure weight. Removing \(\mathrm{AuditAge}\) decreases the overall score from \(0.821\) to \(0.758\), while removing \(\mathrm{ClosedImpact}\) reduces it to \(0.793\). Both variants also increase total token consumption, showing that verification staleness and closed downstream influence jointly improve audit-point selection. Detailed results and analysis are provided in the \textbf{technical supplement}.

\section{Conclusion}


In this paper, we presented BRA-Audit, a cost-aware runtime defense for LLM MAS. It models MAS communication as a dependency graph and formulates auditing as cumulative-exposure audit point placement. By combining audit gaps with topological impact, BRA-Audit selects limited audit points in long-unaudited and influential regions, shortens audit contexts, limits unchecked propagation, and localizes rollback. Experimental results on structured coordination, complex reasoning, and open-ended tasks support a favorable balance between task recovery and token cost. demonstrates a favorable Pareto-optimal trade-off between task performance and auditing cost.  We believe BRA-Audit can contribute to the development of more trustworthy and cost-aware MAS.

\bibliography{aaai2027}


\end{document}